\documentclass[showpacs,amsmath,amssymb,prb,aps,twocolumn]{revtex4}
\usepackage{graphicx,bm}
\usepackage{bm}

\begin{document}
\title {Lorentz shear modulus of fractional quantum Hall states}
\author {I. V. Tokatly}
\email{ilya_tokatly@ehu.es}
\affiliation{European Theoretical Spectroscopy Facility (ETSF),\\ Departamento de Fisica de Materiales, Universidad del Pais Vasco UPV/EHU, Centro Mixto CSIC-UPV/EHU, 20018 Donostia, San Sebastian, Spain\\
and Moscow Institute of Electronic Technology, Zelenograd, 124498 Russia}
\author{G. Vignale}
\email{vignaleg@missouri.edu}
\affiliation{Department of Physics, University of Missouri-Columbia,
Columbia, Missouri 65211}

\date{\today}
\begin{abstract}
We show that the Lorentz shear modulus of macroscopically homogeneous electronic states in the lowest Landau level  is proportional to the bulk modulus of an equivalent system of interacting classical  particles in the thermodynamic limit. Making use of this correspondence  we calculate the Lorentz shear modulus of Laughlin's fractional quantum Hall states at filling factor $\nu=1/m$ ($m$ an odd integer) and find that it is equal to $\pm \hbar mn/4$, where $n$ is the density of particles and the sign depends on the direction of magnetic field.  This is in agreement with the recent result obtained by Read in arXiv:0805.2507 and corrects our previous result published in Phys. Rev. B {\bf 76}, 161305 (R) (2007).  
\end{abstract}
\maketitle

\section{Introduction}

The response of many-body systems to external fields can be universally formulated in terms of hydrodynamical equations of motion, which follow from the local conservation laws of the number of particles and momentum.
These equations completely describe the dynamics of the particle density $n$ and the particle current density  ${\bf j}$ under the combined action of external and internal (or stress) forces. The most important ingredient of this formulation is the stress tensor $P_{ik}$, which determines the stress force density entering the momentum conservation law: $F_{i}^{\rm stress}=-\partial_{k}P_{ik}$. In general the stress tensor for every particular state of matter is a universal functional of the current density~\cite{VigUllCon1997} (or, equivalently, of the deformation tensor~\cite{TokatlyPRB2005b,TokatlyPRB2007}). In fact, the form of $P_{ik}[{\bf j}]$ encodes all the dynamical information about a given many-body state. 

Unfortunately it is an extremely difficult, if not hopeless task to determine the general form of the stress tensor as a functional of the current. As usual the problem simplifies in the linear response regime as the stress tensor becomes a linear functional of the strain tensor $u_{ik}=\frac{1}{2}(\partial_{i}u_{k}+\partial_{k}u_{i})$, where ${\bf u}({\bf r},\omega)$ is the displacement vector field. The coefficients of this linear functional form the rank-4 tensor of elasticity $Q_{ijkl}(\omega)$, which is, in general, a function of frequency. This tensor describes a stress response to a time dependent deformation of the system. In a macroscopically isotropic two-dimensional (2D)  system subjected to a perpendicular magnetic field the rank-4 tensor $Q_{ijkl}$ contains only three independent components and can be parametrized by three dynamic ``elastic moduli'' $K(\omega)$, $\mu(\omega)$, and $\Lambda(\omega)$~\cite{TokatlyPRB2006a,TokatlyPRB2006b,TokVigPRL2007,TokVigPRB2007}.
The moduli $K(\omega)$ and $\mu(\omega)$ describe the response to a local change in volume and to a local volume-preserving (shear) deformation, respectively. Therefore they correspond to the standard bulk and shear moduli of classical elasticity theory.~\cite{LandauVII:e}  The third modulus, $\Lambda(\omega)$, appears only in the presence of the magnetic field and plays an important role in the dynamics of 2D electrons at high magnetic field. This modulus controls the magnitude of a stress proportional to the {\it rate} of volume-preserving deformations, and in this respect is similar to a viscous stress.  However, the corresponding force always acts in a direction perpendicular to the stream lines and is purely nondissipative, like the usual Lorentz force. To underline its nondissipative character, in Refs.~\onlinecite{TokatlyPRB2006b,TokVigPRL2007,TokVigPRB2007} we named the modulus $\Lambda(\omega)$  {\it ``Lorentz shear modulus"}. The same quantity is also known in the literature as ``asymmetric viscosity"~\cite{AvrSeiZog1995}, or ``Hall viscosity"~\cite{Read2008arxiv}. 

The  appearance of a ``nondissipative viscosity'' in the presence of a magnetic field is well known in plasma physics.  The calculation of the corresponding kinetic coefficient $\Lambda_0 = -\lim_{\omega\to 0}\Lambda(\omega)$ for a {\it classical} plasma can be found, for example, in Ref.~\onlinecite{LandauX:e}. The result is
\begin{equation}\label{lambda_classical}
\Lambda_0 = \pm \frac{\hbar n}{2} \frac{k_BT}{\hbar \omega_c}\,,
\end{equation}
 where $T$ is the temperature and $\omega_c$ is the cyclotron frequency (the sign is determined by the direction of the magnetic field). 
 
 In this paper we focus on the microscopic calculation of  $\Lambda_0$ for an extended, macroscopically homogeneous 2D electronic system in the quantum Hall regime, i.e. when all the electrons reside in the lowest Landau level.  For a 2D noninteracting electron liquid in a completely filled Landau level, this calculation was first done by Avron  {\it et al.}  in Ref.~\onlinecite{AvrSeiZog1995}.  These authors observed  that the value of $\Lambda_0$ is proportional to a Berry curvature related to adiabatic changes of geometry, and used this fact to explicitly calculate the modulus.  The result was $\Lambda_0=\pm\hbar n/4$,\cite{note1} which, incidentally, agrees with the classical formula~(\ref{lambda_classical}) if one replaces the thermal  kinetic energy per particle $k_BT$ with the kinetic energy per particle of the lowest Landau level $\hbar \omega_c/2$.
 
 In Ref.~\onlinecite{TokatlyPRB2006a} one of us derived hydrodynamics equations for quantum Hall states at fractional filling factor using a fermionic Chern-Simons theory~\cite{LopFra1991,LopFra1993,SimHal1993,CompositeFermions} at the RPA level. Within this approximation $\Lambda_0$ exactly coincides with the noninteracting result of Ref.~\onlinecite{AvrSeiZog1995}, as reported above, which is not surprising since the composite fermions do not interact in RPA. In a recent paper~\cite{TokVigPRB2007} we applied a formally exact linear response theory to the calculation of the Lorentz shear modulus for strongly correlated quantum Hall states at fractional filling factors (Laughlin states). We showed that $\Lambda(\omega)$ can be expressed in terms of a particular stress-stress correlation function which, in the limit $\omega\to 0$, reduces to the Berry curvature expression of Avron {\it et al.}.  After publication of our work~\cite{TokVigPRB2007} the problem of calculating $\Lambda_0$ was reconsidered by Read~\cite{Read2008arxiv} who noticed an error at the very end of our calculation for Laughlin states.  Making use of the Laughlin plasma analogy~\cite{Laughlin1983} Read correctly calculated the Hall viscosity (Lorentz shear modulus) for Laughlin states at $\nu=1/m$ ($m$ is an odd integer) and also for states with trial wave functions in the form of conformal blocks of a conformal field theory.  For Laughlin states, Read's result is 
 \begin{equation}\label{lambda_Laughlin}
 \Lambda_0 = \pm \frac{\hbar n m}{4}\,, 
 \end{equation}
 whereas in our paper we had incorrectly obtained  $\Lambda_0 = \pm \frac{\hbar n }{4}$.

Read's paper is difficult and contains much more than just the calculation of the Lorentz shear modulus.  We have found that it is possible to give a more elementary derivation of Read's result for filling factors $\nu=1/m$.  The crucial step in the derivation is the recognition that the calculation of the Lorentz shear modulus for a macroscopically homogeneous state of electrons in the lowest Landau level can be mapped to the calculation of the {\it bulk modulus} of an equivalent system of interacting classical particles.  While in general the many-body interactions in this equivalent classical system are prohibitively complicated, they simplify dramatically for Laughlin's quantum Hall states, where one recovers the well-known classical plasma analogy.  This allows a simple calculation of the Lorentz shear modulus.   
The quantum-classical correspondence can also be used ``in reverse".  Namely, from the knowledge of the shear modulus associated with a certain wave function in the lowest Landau level one can in principle obtain the bulk modulus of the equivalent classical system, even if the latter is very complicated.

The structure of this paper is as follows. In Sec.~IIA we introduce the formal definition of the Lorentz shear modulus an discuss its physical significance. Here we also address the delicate question of the correct boundary conditions for the trial many-body wave function to be used in the calculation of the Berry curvature in the thermodynamic limit. The calculation of the Lorentz shear modulus for the Laughlin states is presented in Sec~IIB. In Sec.~IIC we generalize our procedure to other macroscopically homogeneous states in the lowest Landau level. In Sec.~III we summarize our main results and discuss how the new value of $\Lambda_0$ affects our previous analysis of the collective modes of the electron liquid in the fractional quantum Hall regime. 

\section{Lorentz shear modulus in a 2D magnetized electron gas}

\subsection{Definitions and physical significance of the Lorentz shear modulus}

Let us consider a 2D electron gas confined to the $(x,y)$ plane and subjected to a perpendicular magnetic field ${\bf B}=B\hat{\bf z}$. In the linear response regime and in the long wavelength limit the exact stress tensor $P_{ij}$ takes the following ``elastic'' form \cite{TokVigPRL2007,TokVigPRB2007}
\begin{equation}
 P_{ij}({\bf r},\omega) = - Q_{ijkl}(\omega)u_{kl}({\bf r},\omega),
\label{hook_law1}
\end{equation}
where $u_{kl}=\frac{1}{2}(\partial_{k}u_{l}+\partial_{l}u_{k})$ is the strain tensor, and ${\bf u}$ is the displacement vector defined in the standard way, namely  $\partial_t{\bf u} = {\bf j}/n= {\bf v}$  is  the velocity field of the electron liquid. The frequency dependent coefficients $Q_{ijkl}(\omega)$ in the linear functional of Eq.~(\ref{hook_law1}) form the dynamic tensor of elasticity. The general structure of this tensor is essentially fixed by symmetry. In particular, for a macroscopically isotropic state the rank-4 tensor $Q_{ijkl}$ is completely determined by only three independent ``elastic moduli'' 
\begin{eqnarray}\nonumber
  Q_{ijkl}(\omega) &=& K(\omega)\delta_{ij}\delta_{kl} + \mu(\omega)(\delta_{ik}\delta_{jl} + \delta_{il}\delta_{jk} - \delta_{ij}\delta_{kl}) \\
&+& i\omega\frac{\Lambda(\omega)}{2}(\varepsilon_{ik}\delta_{jl}+\varepsilon_{jk}\delta_{il}+
\varepsilon_{il}\delta_{jk}+\varepsilon_{jl}\delta_{ik})
\label{Q}
\end{eqnarray}
Inserting Eq.~(\ref{hook_law1}) into Eq.~(\ref{Q}) we get the folowing representation for the stress tensor for any isotropic state of a 2D magnetized electron gas
\begin{eqnarray}
 \nonumber
P_{ij} = &-& K\delta_{ij}u_{kk} - \mu (2u_{ij}-\delta_{ij}u_{kk})\\
 &+& \Lambda (\varepsilon_{ik}v_{kj} +\varepsilon_{jk}v_{ki}),
\label{hook_law2}
\end{eqnarray}
where $v_{ij}=-i\omega u_{ij} = \frac{1}{2}(\partial_i v_j + \partial_j v_i)$ is the rate of change of the strain. The first two terms in Eq.~(\ref{hook_law2}) describe the stress arising from a local change in volume ($u_{kk}=\nabla{\bf u}$) and from a volume-preserving (traceless, or shear) deformation ($2u_{ij}-\delta_{ij}u_{kk}$), respectively. Hence the corresponding coefficients $K(\omega)$ and $\mu(\omega)$ have the same meaning as the standard bulk and shear moduli of a homogeneous elastic medium.~\cite{LandauVII:e}  The third term in Eq.~(\ref{hook_law2}) is proportional to the rate of change of the strain, i.e., to velocity gradients, which looks somewhat similar to a viscous stress. However, for real $\Lambda$ this term is time-reversal invariant and, therefore, does not cause dissipation. To get a more intuitive picture we consider a ``shear flow'' with the velocity distribution of the form ${\bf v} = (v_x(y), 0)$ (for example, two oppositely directed streamlines). In this simple case the $\Lambda$-part of the stress tensor has only two nonzero elements, 
\begin{equation}\label{LambdaStresses}
P_{xx}^{\Lambda}=-P_{yy}^{\Lambda} = \Lambda\partial_y v_x(y).
\end{equation}
This distribution of stresses corresponds to forces, exerted on the faces of an infinitesimal rectangle, which squeeze or repel (depending on the sign of $\Lambda$) two opposite stream lines. The important point is that these forces are always perpendicular to the local direction of the streamlines, as shown in Fig.~1.  Hence the effect of the third term in Eq.~(\ref{hook_law2}) is analogous to that of the Lorentz (Ampere) force acting between two currents, and causes no dissipation.  Having in mind this physical picture we prefer to call the coefficient $\Lambda(\omega)$ the {\it ``Lorentz shear modulus"} in contrast to more formal terms such as  ``asymmetric viscosity''~\cite{AvrSeiZog1995} or ``Hall viscosity''~\cite{Read2008arxiv}. 

\begin{figure}[]
\includegraphics[width=3.0in]{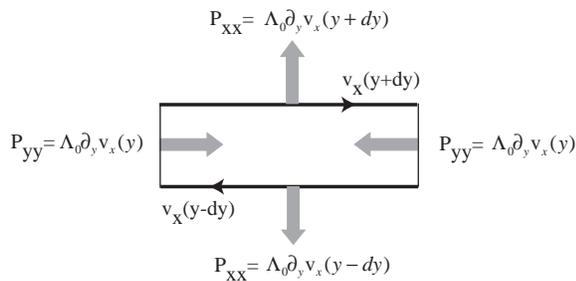}
\caption{Lorentz shear forces acting on a small rectangular element of the electron fluid according to Eq.~(\ref{LambdaStresses}).} 
\label{Fig1}
\end{figure}

\subsection{Linear response approach to the microscopic calculation of the Lorentz shear modulus}

Let us now address the problem of the microscopic calculation of the Lorentz shear modulus in the limit of zero frequency. In our recent paper \cite{TokVigPRB2007} we expressed $\Lambda_0 \equiv -\lim_{\omega\to 0}\Lambda(\omega)$ in terms of the stress-stress correlation function in the following manner:
\begin{equation}
 \Lambda_0 = - \lim_{\omega\to 0}\lim_{{\bf q}\to 0}\frac{1}{\omega}
{\rm Im}\langle\langle{\hat P_{xx}};{\hat P_{xy}}\rangle\rangle_{{\bf q},\omega}\,.
\label{Lambda_def}
\end{equation}
The stress tensor operator ${\hat P_{ij}}$ is formally defined as follows (see, e.~g., Ref.~\onlinecite{TokatlyPRB2007})
\begin{equation}\nonumber
 \hat P_{ij}({\bf r},t)  = -2\left[\frac{\delta {\hat H}[g_{ij}]}{\delta g_{ij}({\bf r},t)}\right]_{g_{ij}=\delta_{ij}},
\end{equation}
where ${\hat H}[g_{ij}]$ is the Hamiltonian of the system in a ``deformed'' space with metric $g_{ij}({\bf r},t)$.   The subscript ${\bf q}$ in Eq.~(\ref{Lambda_def}) means that we are actually considering the correlation function for  the Fourier component  of $ \hat P_{ij}({\bf r},t)$ at wave vector ${\bf q}$, where ${\bf q}$ tends to zero before $\omega$.   It is convenient to parametrize the metric tensor as follows \cite{AvrSeiZog1995,Levay1995,TokVigPRB2007}
\begin{equation}
g_{ij} = \frac{J}{\tau_2}\left(
\begin{array}{cc}
  1     &   \tau_1 \\ 
 \tau_1 &   |\tau|^2
\end{array}
 \right),
\label{metric}
\end{equation}
where $\tau = \tau_1 + i\tau_2$.  A picture of the deformation of the Euclidean plane corresponding to this choice of metrics is shown in Fig.~2. Making use of the Lehmann representation~\cite{GV05} for the stress-stress correlation function and performing the standard manipulations one can transform Eq.~(\ref{Lambda_def}) to the following Berry curvature form
\begin{equation}
\Lambda_0 = \frac{2\hbar}{L^2}{\rm Im}\left.\left\langle\frac{\partial\Psi_0}{\partial\tau_1}\right\vert\frac{\partial\Psi_0}{\partial\tau_2} \right\rangle
 \label{Lambda_Berry}
\end{equation}
where $L^2$ is the area of the system, $\Psi_0$ is the ground state wave function in a {\it homogeneously} deformed space with a constant metric of Eq.~(\ref{metric}), and the $\tau$-derivatives are calculated at $\tau_1=0$, $\tau_2=1$, and $J=1$.

\begin{figure}[]
\includegraphics[width=2.5in]{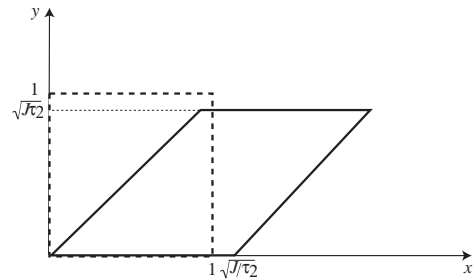}
\caption{Deformation of the Euclidean plane corresponding to the metric of Eq.~(\ref{metric}).  The square is transformed into a parallelogram.  The slope of the oblique side coincides with the direction of $\tau$ in the complex plane.  We have chosen $J=1$, $\tau_1=\tau_2<1$.} 
\label{Fig2}
\end{figure}

Equation (\ref{Lambda_Berry}) was first derived in Ref.\onlinecite{AvrSeiZog1995} from the adiabatic response theory. Our linear response derivation shows that there is a delicate point in the identification of the Berry curvature [the right hand side in Eq.~(\ref{Lambda_Berry})] with the physical Lorentz shear modulus.  Physically such a modulus describes a stress response to a deformation whose wavelength is much larger than any internal scale of the system (e.g. the interparticle distance, the correlation length, etc.), but still much smaller than the size of the sample $L$. At the level of the linear response formula, Eq.~(\ref{Lambda_def}), this means that the thermodynamic limit should be performed before the limit ${\bf q}\to 0$.  This guarantees that the calculated  elastic modulus does not depend on the geometry of the sample (provided the sample is sufficiently large), i.e. it is a bulk property of the ``material", as it should. However, in the derivation of the Berry curvature formula the order of limits was explicitly interchanged. The right hand side of Eq.~(\ref{Lambda_Berry}) contains the ground state wave function of a finite system with a finite number of particles $N$. Physically this describes the response to a homogeneous deformation of the whole sample which in general may depend on the sample geometry.\cite{note2}   Nonetheless it is still possible to obtain the correct physical result  even with the interchanged order of limits. The key is to use boundary conditions for the wave function $\Psi_0$, which are compatible with the symmetry of the physical plane wave perturbation (the deformation in our particular case).  For, in this way we ensure that the ${\bf q} \to 0$ limit of the deformed state goes smoothly to the homogeneous  deformation of the ground-state in the same geometry.
In practice this simply means that the correct result is guaranteed if one studies a sample of rectangular shape, which satisfies periodic boundary conditions either in two directions (torus)  or in one direction (cylinder).  On the other hand a sample of circular shape would not be compatible with the symmetry of the plane wave, because the presence of a finite wave vector, no matter how small, always breaks rotational symmetry.\cite{note3}

The choice of the torus geometry in the present context may also cause some technical problems which require a special care. On a torus the ground state for gapped quantum Hall states is degenerate. This nicely demonstrates the topological nature of these states, but at the same time introduces an artificial degeneracy which is not present in any real physical sample. Clearly, this degeneracy should not affect the physical elastic moduli.  To guarantee that this is indeed the case we put our system of $N$ electrons on a cylinder with a circumference $L$ in such a way that it occupies the area $L^2= N_S (2\pi \ell^2)$, where $N_S=N/\nu$ is the number of flux quanta (one flux quantum = $hc/e$), and $\ell = \sqrt{\hbar c/eB}$ is the magnetic length. The thermodynamic limit which we take at the end of calculations corresponds to $N\to\infty$ at fixed filling factor $\nu=N/N_S$.

For the Hamiltonian defined on a space with the metric of Eq.~(\ref{metric}) the $N$-body wave function which lies entirely in the lowest Landau level and satisfies periodic boundary condition in the $x$-direction can be written in the following general form
\begin{equation}
 \Psi_0({\bf r}_1,\dots,{\bf r}_N)=Z^{-\frac{1}{2}}(\tau_1,\tau_2)f(\eta_1,\dots,\eta_N)
\prod_{i=1}^{N}e^{\frac{i}{2\ell^2}\tau y_i^2}
\label{Psi}
\end{equation}
where $Z^{-\frac{1}{2}}$ is the normalization factor, $\eta_j\equiv\exp(2\pi i\frac{x_j + \tau y_j}{L})$, and $f(\eta_1,\dots,\eta_N)$ is an analytic function of its arguments. A crucial observation is that $\Psi_0$, Eq.~(\ref{Psi}), apart from the normalization factor, is an analytic function of the complex variable $\tau$. This enables us to express the Berry curvature solely in terms of the normalization factor:\cite{Levay1995,TokVigPRB2007}
\begin{equation}
 2{\rm Im}\left.\left\langle\frac{\partial\Psi_0}{\partial\tau_1}\right\vert\frac{\partial\Psi_0}{\partial\tau_2} \right\rangle
= \frac{1}{2}\left(\frac{\partial^2}{\partial\tau_1^2} + \frac{\partial^2}{\partial\tau_2^2} \right)\ln Z\,.
\label{Berry}
\end{equation}
Hence the problem of microscopic calculation of the Lorentz shear modulus reduces to the calculation of the normalization factor. For our particular choice of a quantum Hall system on a cylinder the problem simplifies even further since the normalization factor depends only on $\tau_2$. Indeed, by shifting all the $x$-variables, $(x_j+\tau_1 y_j)\to x_j$, and rescaling all $y$-variables, $\tau_2 y_j\to y_j$, we reduce the normalization integral to the following form
\begin{eqnarray}\nonumber
 Z(\tau_2) &=& \tau_2^{-N}\prod_{k=1}^N\int\limits_0^L dx_k\int\limits_{-\infty}^{\infty}dy_k \\
&\times& \left\vert\left(e^{2\pi iz_1/L},\dots,e^{2\pi iz_N/L} \right)\right\vert^2 e^{-\sum_{j=1}^{N}\frac{y_j^2}{\tau_2 \ell^2}}\,,\nonumber\\
\label{Z}
\end{eqnarray}
where $z_k = x_k + iy_k$. Hence the final formula for the Lorentz shear modulus in the state $\Psi_0$ takes the form
\begin{equation}
 \Lambda_0 = \frac{\hbar}{2L^2}\left[\frac{\partial^2}{\partial\tau_2^2}\ln Z(\tau_2)\right]_{\tau_2 =1},
\label{Lambda_Z}
\end{equation}
with $Z(\tau_2)$ defined after Eq.~(\ref{Z}). In the next subsections we calculate this integral using the Laughlin classical plasma analogy.

\subsection{Calculation of the Lorentz shear modulus for Laughlin states}

The cylindrical generalization of the Laughlin trial function at $\nu=1/m$ contains the analytic factor $f$ of the following form \cite{Thouless1984} 
\begin{equation}
 f\left(z_1,\dots,z_N \right) = \prod_{j<k}\left(e^{2\pi iz_j/L} - e^{2\pi iz_k/L} \right)^m\,.
\label{f}
\end{equation}
Inserting this equation into Eq.~(\ref{Z}) we represent the normalization integral in a form of a partion function of $N$ classical particles,
\begin{equation}
 Z(\tau_2) = \tau_2^{-N}\prod_{k=1}^N\int\limits_0^L dx_k\int\limits_{-\infty}^{\infty}dy_k
 e^{-\frac{1}{\nu} W({\bf r}_1,\dots,{\bf r}_N)},
\label{Z_m}
\end{equation}
at the ``temperature'' $\nu=1/m$ and with the following energy
\begin{equation}
 W({\bf r}_1,\dots,{\bf r}_N) = \frac{2\pi N}{\tau_2 L^2}\sum_{j=1}^N y_j^2 + 
\frac{1}{2}\sum_{j\ne k}\tilde{V}({\bf r}_j,{\bf r}_k),
\label{W}
\end{equation}
where we used the identity $1/m=N/N_S\equiv 2\pi N \ell^2/L^2$, and introduced the notation
\begin{equation}
 \label{tildeV}
\tilde{V}({\bf r},{\bf r}') = -\ln\left|e^{2\pi iz_j/L} - e^{2\pi iz_k/L} \right|^2.
\end{equation}
The first term in Eq.~(\ref{W}) has a clear physical interpretation: it is the Coulomb potential of a homogeneously distributed (on the cylindrical surface) {\it positive} charge with density $\rho = N/(\tau_2 L^2)$. Note that after the rescaling of the $y$-coordinate the size of the system along the axis of the cylinder becomes $L_y=\tau_2 L$: the classical system with the probability distribution $\sim\exp[-mW(\{{\bf r}_j\})]$ occupy the region with $0<y<\tau_2 L$. Hence the quantity $N/(\tau_2 L^2)$ entering the first term in Eq.~(\ref{W}) exactly coinsides with the physical density of this classical system, in perfect agreement with the idea of the Laughlin classical plasma analogy.~\cite{Laughlin1983}  However, at variance with the original circularly symmetric Laughlin wave function, the second term in Eq.~(\ref{W}) does not look like the interaction energy of a Coulomb plasma. To show that this is nonetheless the case, we express the two-point ``potential'' $\tilde{V}({\bf r},{\bf r}')$, Eq.~(\ref{tildeV}), in terms of the physical Coulomb interaction $V({\bf r}-{\bf r}')$ between two point particles on a cylinder, (see Appendix~A)
\begin{equation}
 \label{p-pInteraction1}
V({\bf r}-{\bf r}') = -2\pi\frac{|y-y'|}{L} - \ln\left| 1 - e^{2\pi i\frac{x-x' +i|y-y'|}{L}} \right|^2.
\end{equation}
The identity of Eq.~(\ref{p-pInteractionA2}) allows us to relate $\tilde{V}({\bf r},{\bf r}')$ to $V({\bf r}-{\bf r}')$
\begin{equation}
 \label{p-pInteraction2}
\tilde{V}({\bf r},{\bf r}') = 2\pi\frac{y+y'}{L} + V({\bf r}-{\bf r}').
\end{equation}
Substituting this equation in to Eq.~(\ref{W}) we transform the energy $W(\{{\bf r}_j\})$ to the folowing form
\begin{equation}
\label{W1}
 W(\{{\bf r}_j\}) = -\tau_2\frac{\pi}{2}(N-1)^2 + \frac{2\pi N}{\tau_2 L^2}\sum_{j=1}^N \tilde y_j^2 +
\frac{1}{2}\sum_{j\ne k}V(\tilde{\bf r}_j-\tilde{\bf r}_k),
\end{equation}
where $\tilde x_k=x_k$ and $\tilde y_k = y_k +\frac{1}{2}\tau_2 L(1-\frac{1}{N})$. Hence the main effect of the difference between $\tilde{V}({\bf r},{\bf r}')$ and $V({\bf r}-{\bf r}')$ is a shift of the $y$-coordinate by a half of the system size $\sim \tau_2 L/2$. In what follows we remove this shift by moving the origin to the center of the slab: $\tilde y_k\to y_k$. 

Thus the energy $W(\{{\bf r}_j\})$ exactly corresponds to the energy of a system of $N$ classical charges on an infinite homogeneously charged cylinder. To proceed further in the calculation of the partition function of this system we separate a ``neutralizing'' part of the background with density
\begin{equation}
 \label{n_back}
n_b({\bf r}) = \frac{N}{\tau_2 L^2}\theta\left(\tau_2\frac{L}{2} - |y| \right),
\end{equation}
and total charge $\int d{\bf r}n_b({\bf r})=N$. The next step is to separate the energy of interaction of the particles with $n_b({\bf r})$ from the total energy of Eq.~(\ref{W1}). Using Eqs.~(\ref{Upb}) and (\ref{Wbb}) of Appendix~A we find the following representation for $W(\{{\bf r}_j\})$
\begin{equation}
 \label{W2}
W = -\tau_2\pi\left(\frac{2N^2}{3}-N+\frac{1}{2}\right) + E_{\rm{NJ}}(\{{\bf r}_j\})+E_{\rm{S}}(\{{\bf r}_j\}).
\end{equation}
The second term, $E_{\rm{NJ}}(\{{\bf r}_j\})$, in this equation is the energy of a neutral slab of jellium,
\begin{equation}
 \label{E_NJ}
E_{\rm{NJ}} = \frac{1}{2}\int d{\bf r}d{\bf r}'V({\bf r}-{\bf r}') 
[\Delta\hat \rho({\bf r})\Delta\hat\rho({\bf r}') - \delta({\bf r}-{\bf r}')\hat\rho({\bf r})]
\end{equation}
where $\hat\rho({\bf r})$ is the microscopic {\it particle density}, and $\Delta\hat\rho({\bf r})$ is the microscopic {\it charge density} of jellium:
\begin{equation}
 \label{density}
\hat\rho({\bf r})=\sum_{j=1}^N\delta({\bf r}-{\bf r}_j), \quad 
\Delta\hat\rho({\bf r})=\hat\rho({\bf r}) - n_b({\bf r}).
\end{equation}
The third term in Eq.~(\ref{W2}) corresponds to the surface (edge) contribution 
\begin{equation}
 \label{E_S}
E_{\rm S} = \frac{2\pi N}{\tau_2 L^2}\sum_{j=1}^N \left(|y_j| - \frac{\tau_2 L}{2} \right)^2
\theta\left(|y_j| - \frac{\tau_2 L}{2} \right),
\end{equation}
which is the potential energy of particles outside the ``neutralizing'' part of the background (see Fig.~3).

\begin{figure}[]
\includegraphics[width=3.0in]{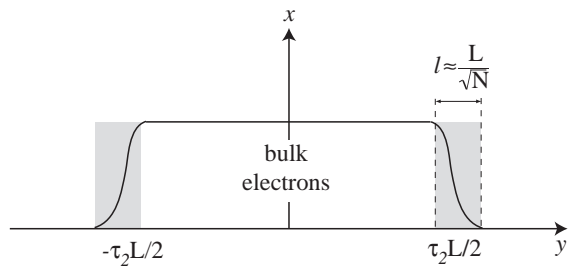}
\caption{Distribution of the electron density in the $y$ direction.  Notice the two small edge regions, whose width is proportional to the mean interparticle distance $\ell \sim \frac{L}{\sqrt{N}}$.} 
\label{Fig3}
\end{figure}

Eq.~(\ref{W2}) is extremely useful for the analysis of the relevant limit $N\to\infty$ of the partition function. It it physically obvious that the contribution, Eq.~(\ref{E_S}), should be irrelevant in the thermodynamic limit as it involves a small fraction of particles in the edge region.  Indeed, from Fig.~3 we see that the contribution to $E_{\rm S}$ comes from a region whose width is of the order of the mean interparticle distance $\ell\sim\frac{1}{\sqrt{n}}=\frac{L}{\sqrt{N}}$ (since it is proportional to the magnetic length we use the same notation). Then a simple, order of magnitude estimate of $E_{\rm S}$ is
\begin{equation}
E_S \sim \frac{N}{L^2}(L\ell n) \ell^2 \sim \sqrt{N}\,.
\end{equation}
where $L\ell n\sim\sqrt{N}$ is the average number of particles in the edge region.
This shows that in the limit $N\to\infty$ $E_{\rm S}$ grows as $\sqrt{N}$ and is negligible in comparison with the first and the second terms in Eq.~(\ref{W2}), which are proportional to $N^2$ and $N$ respectively. Inserting the energy of Eq.~(\ref{W2}) into Eq.~(\ref{Z_m}) and neglecting $E_{\rm S}$, we rewrite the normalization integral as follows
\begin{equation}
 \label{Z_m1}
Z(\tau_2) = L^{2N}e^{m\tau_2\pi\left(\frac{2N^2}{3}-N+\frac{1}{2}\right)}\frac{Z_{\rm{NJ}}}{Z_0},
\end{equation}
where $Z_{\rm{NJ}}=\langle \exp[-mE_{\rm{NJ}}(\{{\bf r}_j\})]\rangle$ is the partition function of the slab of jellium, and $Z_0 = (\tau_2 L^2)^N$ is the partition function of $N$ noninteracting particles in the slab. Hence Eq.~(\ref{Z_m1}) can be equivalently represented in the form
\begin{equation}
 \label{Z_m2}
Z(\tau_2) = L^{2N}e^{m\tau_2\pi\left(\frac{2N^2}{3}-N+\frac{1}{2}\right)}e^{-mF_{\rm{int}}(\tau_2)},
\end{equation}
where $F_{\rm{int}} = -\frac{1}{m}\ln(Z_{\rm{NJ}}/Z_0)$ is the interaction free energy of the classical jellium model. The final step is to substitute Eq.~(\ref{Z_m2}) into the expression of Eq.~(\ref{Lambda_Z}) for the Lorentz shear modulus. The first, non-extensive exponent ($\sim N^2$) in Eq.~(\ref{Z_m2}) does not contribute to $\Lambda_0$ since it is linear in $\tau_2$. Therefore Eq.~(\ref{Lambda_Z}) reduces to the form
\begin{equation}
 \Lambda_0 = -\frac{\hbar m}{2}\left[\frac{\partial^2}{\partial\tau_2^2}
\tau_2 \frac{F_{\rm{int}}(\tau_2)}{\tau_2 L^2}\right]_{\tau_2 =1}.
\label{Lambda_F}
\end{equation}
Now we can easily perform the thermodynamic limit. In this limit, the free energy density of a neutral jellium depends only on the density of particles, which is equal to the background charge density $\rho$ of the jellium:  $F_{\rm{NJ}}/(\tau_2 L^2)=f_{\rm{NJ}}(\rho)$, where $\rho=n/\tau_2$ and $n=N/L^2$ is the density of electrons in the original quantum Hall system. Thus in the thermodynamic limit Eq.~(\ref{Lambda_F}) simplifies as follows
\begin{equation}
 \Lambda_0 = -\frac{\hbar m}{2}\left[\frac{\partial^2}{\partial\tau_2^2}\tau_2 
f_{\rm{int}}\left(\frac{n}{\tau_2}\right)\right]_{\tau_2 =1},
\label{Lambda_f}
\end{equation}
where $f_{\rm{int}}(\rho)$ is the interaction free energy density of a 2D classical jellium model with logarithmic interaction between the particles. The key observation is that $\tau_2$ enters Eq.~(\ref{Lambda_f}) as an effective volume (the area in 2D). Hence the second derivative of the energy with respect to $\tau_2$ in Eq.~(\ref{Lambda_f}) is nothing but the isothermal bulk modulus:
\begin{equation}
 \Lambda_0 = \frac{\hbar m}{2}\left[\frac{\partial}{\partial\tau_2} 
P_{\rm{int}}\left(\frac{n}{\tau_2}\right)\right]_{\tau_2 =1} =-\frac{\hbar m}{2}K_{\rm{int}}(n),
\label{Lambda_K}
\end{equation}
where $P_{\rm{int}}(\rho)$, and $K_{\rm{int}}(\rho)$ are the interaction contributions to the pressure and bulk modulus, respectively. Equation (\ref{Lambda_K}) is the main result of the present paper. It shows that the calculation of the Lorentz shear modulus in a quantum Hall system reduces to the calculation of the bulk modulus of an equivalent classical system. In the next subsection we will demonstrate that this result holds more generally for any macroscopically homogeneous quantum Hall state. 

For the Laughlin states considered in this subsection we need the bulk modulus $K_{\rm{int}}(\rho)$ of a 2D Coulomb plasma with logarithmic interaction, which can be calculated exactly.\cite{Caillol82} The easiest way is to use the virial representation for the interaction pressure
\begin{equation}
 \label{virial}
P_{\rm{int}} = -\frac{1}{4S}\int d{\bf r}d{\bf r}'({\bf r}-{\bf r}')
\frac{\partial V({\bf r}-{\bf r}')}{\partial ({\bf r}-{\bf r}')}\rho({\bf r})\rho({\bf r}')
[g({\bf r},{\bf r}')-1]
\end{equation}
where $S$ is the area of the system and $g({\bf r},{\bf r}')$ is the pair correlation function. Inserting $V({\bf r})=-\ln|{\bf r}|^2$ into Eq.~(\ref{virial}) and using the sum rule
$$
\int d{\bf r}'\rho({\bf r}')[g({\bf r},{\bf r}')-1] =-1,
$$
we obtain the interaction pressure and the corresponding bulk modulus in the form
$$
P_{\rm{int}}(\rho) = K_{\rm{int}}(\rho) = -\frac{1}{2}\rho.
$$
This implies the following final result for the Lorenz shear modulus of the Laughlin quantum Hall liquids
\begin{equation}
 \label{Lambda_fin}
\Lambda_0=\frac{\hbar nm}{4}  = \frac{\pi\hbar}{2{\ell}^2}.
\end{equation}
This equation reproduces the recent result of Read~\cite{Read2008arxiv}. In the case of a full Landau level ($m=1$) it recovers the formula for the ``asymmetric viscosity'' obtained by Avron, et.~al. in Ref.~\cite{AvrSeiZog1995}. It is interesting to note a surprising feature of Eq.~(\ref{Lambda_fin}) : because the density $n$ is proportional to $1/m$ the Lorentz shear modulus for Laughlin states does not depend on the filling factor.

\subsection{Generalization to arbitrary macroscopically homogeneous quantum Hall states}

The result of Eq.~(\ref{Lambda_K}) can be straightforwardly generalized to other macroscopically homogeneous states in the lowest Landau level. These are states in which the macroscopic particle density is uniform.  For example a Wigner crystal, while microscopically inhomogeneous, can be considered homogeneous on a macroscopic scale:  the average density is uniform. In fact, most states of interest in condensed matter physics have this property.  

In general the normalization integral for any wave function $\Psi_0(\{{\bf r}_j\})$ that satisfies cylindrical boundary conditions in the lowest Landau level is representable in the form of Eq.~(\ref{Z_m}), where
\begin{equation}
 W(\{{\bf r}_j\}) = \frac{2\pi N}{\tau_2 L^2}\sum_{j=1}^N y_j^2 + 
\tilde{U}(\{{\bf r}_j\}),
\label{W_general}
\end{equation}
and $\tilde{U}(\{{\bf r}_j\})=-\nu\ln|f(\{{\bf r}_j\})|^2$ is the interaction energy of the equivalent classical system [here $f(\{{\bf r}_j\})$ is the analytic factor in the many body wave function, Eq.~(\ref{Psi})]. The interpretation of the first term in Eq.~(\ref{W_general}) is exactly the same as for the Laughlin states:  it is the energy of classical particles in the electrostatic field of a homogeneous charged background with the density $\rho=N/(\tau_2 L^2)$. Depending on the form of the wave function $\Psi_0$, the second term can contain both two-particle and multi-particle interactions. In general the form of these interactions is extremely complicated: they are neither translationally nor rotationally invariant.  However, if the physical density distribution of the quantum state under consideration is macroscopically homogeneous with a mean density $\bar n=N/L^2$, then the interaction energy $\tilde{U}(\{{\bf r}_j\})$ of the equivalent classical system {\it must} contain a long-range Coulomb contribution of the form $\sum_{j<k}\tilde{V}({\bf r}_j,{\bf r}_k)$. The presence of this pairwise logarithmic interaction is mandatory to compensate the background potential and protect the macroscopic homogeneity of the classical gas. Hence for macroscopically homogeneous states in the lowest Landau level the energy of the equivalent classical plasma should take the form
\begin{equation}
 W(\{{\bf r}_1\}) = \frac{2\pi N}{\tau_2 L^2}\sum_{j=1}^N y_j^2 + 
\sum_{j< k}\tilde{V}({\bf r}_j,{\bf r}_k)+
\tilde{U}_{{\rm sr}}(\{{\bf r}_j\}),
\label{W_general1}
\end{equation}
where the last term $\tilde{U}_{{\rm sr}}(\{{\bf r}_j\})$ can contain only short range (possibly multiparticle, anisotropic, etc.) interactions. Since short range interactions do not spoil the extensive character of the free energy, the chain of arguments that led us from Eq.~(\ref{Z_m}) to Eq.~(\ref{Lambda_f}) remains valid for any macroscopically homogeneous quantum Hall state. Therefore the Lorentz shear modulus for such states can be calculated as follows
\begin{equation}
 \Lambda_0 = \frac{\hbar}{2\nu}\left[\frac{\partial}{\partial\tau_2} 
P_{\rm{int}}\left(\frac{n}{\tau_2}\right)\right]_{\tau_2 =1} =-\frac{\hbar}{2\nu}K_{\rm{int}}(n),
\label{Lambda_K_general}
\end{equation}
where $P_{\rm{int}}(\rho)$, and $K_{\rm{int}}(\rho)$ are the interaction pressure and the isothermal bulk modulus of the corresponding classical plasma. 


As a simple example consider the wave function proposed by Wexler and Ciftja~\cite{Wexler02} for the nematic liquid crystal state at $\nu=1/3$.  The cylindrical generalization of this wave function  contains an analytic factor $f$ the form
\begin{eqnarray}\label{WexlerWavefunction}
&& f\left(z_1,\dots,z_N \right) = \prod_{j<k}\left(e^{2\pi iz_j/L} - e^{2\pi iz_k/L}\right)\nonumber\\&\times& \left(e^{2\pi i(z_j-\alpha/2)/L} - e^{2\pi i(z_k+\alpha/2)/L}\right)
\nonumber\\&\times& \left(e^{2\pi i(z_j+\alpha/2)/L} - e^{2\pi i(z_k-\alpha/2)/L}\right)\,,
\end{eqnarray}
where $\alpha$ is a complex number, which we set equal to $i a$, with $a$ real and positive.  With this choice the system remains invariant under rotations about the axis of the cylinder.  The parameter $a$ is the microscopic distance (of the order of the magnetic length) by which two of the three zeroes of the  $\nu=1/3$  Laughlin  wave function are displaced from their ``regular" position on top of the particle. The Laughlin wavefunction  is recovered by setting $a=0$.  It is now straightforward to verify that  the classical system that is equivalent to this wave function contains, in addition to the interactions of Eq.~(\ref{W}), a two-body short-range interaction of the form  (in the physically relevant 2D regime, $|z_i-z_j|\ll L$)
\begin{equation}\label{U_shortrange}
\tilde{U}_{{\rm sr}}(\{{\bf r}_j\}) = -\frac{\nu}{2} \sum_{i\neq j}  \ln \left \vert 1+ \frac{a^2}{(z_i-z_j)^2}\right\vert^2.
\end{equation}
This interaction is strongly anisotropic and exhibits  logarithmic singularities at interparticle distance $0$ (independent of direction) and $a$ (along the $y$ axis).  It is clearly very difficult to calculate the bulk modulus of this system from classical statistical mechanics. However, if we could, by some independent method, calculate the Lorentz shear modulus of the wave function~(\ref{WexlerWavefunction}), then the problem would be solved. 

\section{Conclusion}

In this paper we have shown that the calculation of the Lorentz shear modulus in a macroscopically homogeneous electronic system in the lowest Landau level can be mapped to the calculation of the bulk modulus of an equivalent classical system.  Application of this approach to Laughlin's fractional quantum Hall states gives a value of the modulus independent of filling factor in agreement with the result of Ref.~\onlinecite{Read2008arxiv} and in contrast with what we had erroneously found in Ref.~\onlinecite{TokVigPRB2007}. In that paper, the incorrect value of the Lorentz shear modulus was used in the calculation of the collective modes of the fractional quantum Hall liquid, and the resulting dispersions were found to be in good agreement with the ones obtained  in Ref.~\onlinecite{TokVigPRL2007} by a more phenomenological approach. 
 
A serious problem arises when we repeat the calculation of the collective mode dispersions according to Eq.~(21) of Ref.~\onlinecite{TokVigPRB2007}, but using the correct value of  $\Lambda_0$ for the $\nu=1/3$ Laughlin state.  We get an instability  in which the upper collective mode has negative oscillator strength. This means that the simplest single-mode ansatz for $\Lambda(\omega)$, Eq.~(20)  of Ref.~\onlinecite{TokVigPRB2007} fails.  Most likely this indicates that in reality the high-frequency shear modulus $\Lambda(\omega\gg \Delta)$ does not vanish as it was assumed in that equation. Indeed, by adding a frequency independent term to Eq.~(20) of Ref.~\onlinecite{TokVigPRB2007} we can recover all low-${\bf q}$ predictions of Ref.~\onlinecite{TokVigPRL2007} without contradicting the stability criterion.

\section{Acknowledgments} 

One of us (GV) gratefully acknowledges support from  NSF under Grant No. DMR-0705460. The work of IVT was supported by the Ikerbasque Foundation.
We thank Nick Read and Yosi Avron for their help in clarifying the numerical factor of Eq.~(\ref{lambda_Laughlin}) and for bringing to our attention Eq.~(\ref{lambda_classical}).

\appendix
\section{Electrostatics on a cylinder}

In this appendix we calculate a few important ingredients of a classical neutral jellium model on a cylinder

The interaction potential $V({\bf r}-{\bf r}')$ of two point unit charges living on a cylindrical surface is a periodic solution of a 2D Poisson equation
\begin{eqnarray}
 \label{Poisson}
&&\nabla^2 V(x,y) = - 4\pi \delta(x-x'+nL)\delta(y-y'),\\
\nonumber
&& V(x+L,y)=V(x,y),
\end{eqnarray}
where $L$ is the circumference of the cylinder and $n$ is an integer number. This equation is readily solved by the Fourier transformation. The result takes the following form
\begin{equation}
 \label{p-pInteractionA1}
V({\bf r}-{\bf r}') = -2\pi\frac{|y-y'|}{L} - \ln\left| 1 - e^{2\pi i\frac{x-x' +i|y-y'|}{L}} \right|^2
\end{equation}
It is easy to see that this potential has correct 1D and 2D asymptotic forms
\begin{equation}\nonumber
V({\bf r}-{\bf r}') = \left\{ \begin{array}{cc} 
-\frac{2\pi}{L}|y-y'|, & |y-y'|\gg L \\
-\ln |{\bf r}-{\bf r}'|^2, &  |{\bf r}-{\bf r}'|\ll L
\end{array}\right.
\end{equation}

It is also straightforward to check the following useful representation for the potential $V({\bf r}-{\bf r}')$
\begin{equation}
 \label{p-pInteractionA2}
V({\bf r}-{\bf r}') = -2\pi\frac{y+y'}{L} - \ln\left|e^{2\pi iz/L} - e^{2\pi iz'/L} \right|^2.
\end{equation}

Now we find the interaction potential $U_{pb}({\bf r})$ of a negatively charged particle with a positive background charge density $n_b({\bf r})$ that is homogeneously distributed on a cylinder and occupies the region $-\frac{L\tau_2}{2}<y<\frac{L\tau_2}{2}$ 
$$
n_b({\bf r}) = \rho \theta\left(\frac{L\tau_2}{2} - |y|\right),
$$
where $\rho=\frac{N}{LL\tau_2}$, and $N$ is the total charge of the background. The potential $U_{pb}({\bf r})$ is given by the following integral
\begin{eqnarray}\nonumber
&&U_{pb}({\bf r}) = -\int d{\bf r}' V({\bf r}-{\bf r}') n_b({\bf r}') 
= -\pi\frac{N}{2}\frac{L\tau_2}{L} \\
&-& 2\pi\rho y^2 + 2\pi\rho \left(|y| - \frac{L\tau_2}{2} \right)^2\theta\left(|y|-\frac{L\tau_2}{2}\right)
 \label{Upb}
\end{eqnarray}

Similarly one calculates the background-background interaction energy
\begin{equation}
 W_{bb} = \frac{1}{2}\int d{\bf r}d{\bf r}' V({\bf r}-{\bf r}') n_b({\bf r}')n_b({\bf r}) =
\pi\frac{N^2}{3}\frac{L\tau_2}{L}
\label{Wbb}
\end{equation}


\end{document}